\documentclass[aip,
rsi,
 amsmath,amssymb,
 reprint,%
]{revtex4-1}

\usepackage{graphicx}
\usepackage{dcolumn}
\usepackage{bm}
\usepackage{cancel}

\usepackage[utf8]{inputenc}
\usepackage[T1]{fontenc}
\usepackage{mathptmx}
\usepackage{etoolbox}
\usepackage{booktabs}
\usepackage[thinc]{esdiff}
\usepackage[italicdiff]{physics}
\usepackage[version=4]{mhchem}
\usepackage{makecell}
\usepackage[dvipsnames]{xcolor}
\usepackage{tikz}
 
\usepackage{mdframed}
\usepackage{float}
\usepackage{soul}
\usepackage[normalem]{ulem}

\makeatletter
\def\@email#1#2{%
 \endgroup
 \patchcmd{\titleblock@produce}
  {\frontmatter@RRAPformat}
  {\frontmatter@RRAPformat{\produce@RRAP{*#1\href{mailto:#2}{#2}}}\frontmatter@RRAPformat}
  {}{}
}%
\makeatother
\begin{document}


\title[]{Two lock-in amplifiers based $3\omega$ technique: a practical guide\\ for thermal conductivity experiments in bulk samples}
\author{A. Henriques}
\affiliation{Laboratory for Quantum Matter under Extreme Conditions, Institute of Physics, University of São Paulo, Brazil}%
\author{M. Santoma}
\affiliation{Laboratory for Quantum Matter under Extreme Conditions, Institute of Physics, University of São Paulo, Brazil}%
\author{S. Wirth}%
\affiliation{ 
Max-Planck Institute for Chemical Physics of Solids, Dresden, Germany}
\author{J. Larrea Jiménez}
\affiliation{Laboratory for Quantum Matter under Extreme Conditions, Institute of Physics, University of São Paulo, Brazil}%
\author{V. Martelli}
\affiliation{Laboratory for Quantum Matter under Extreme Conditions, Institute of Physics, University of São Paulo, Brazil}%
\email{valentina.martelli@usp.br}

\date{\today}

\begin{abstract}
The accurate determination of thermal conductivity $\kappa(T)$ in bulk materials at room temperature and above is crucial for evaluating their compatibility for specific applications. The 3$\omega$ technique is an established methodology for studying the thermal conductivity of thin films, becoming particularly suitable in the case of bulk specimens for $T\gtrsim$300~K, where standard stationary techniques require significant corrections for radiative losses. Although this method has been employed in several works, it remains not widely adopted because its implementation demands considerable sophistication, including experiment design, thin film deposition techniques, and choices of the geometry of the current/heat transducer, electronics, and analytical treatment of the signals. 
Based on a critical review of the technique's key technical aspects, this work provides practical support for a rapid and user-friendly implementation, from the design phase through to execution and analysis. We release a Python-based graphical user interface that supports a quantitative estimation of the investigated temperature profiles based on the geometrical parameters (width/length) of the deposited transducer (heater/thermometer metal line) before an experiment, guaranteeing an optimal design of the experimental conditions for each given material under scrutiny.    


\end{abstract}

\maketitle


\section{Introduction}



In the early 1910s, the first measurements of AC heat capacity were conducted by Corbino \cite{Corbino1910, Corbino1911}. The possibility for periodic heating in thermal experiments opened the way for designing a new apparatus where the phase and amplitude of the AC temperature profile can allow for a reduced noise level \cite{Hatta1997}. The late sixties saw intense improvement in the methodology \cite{Sullivan1968, Gobrecht1971}. Thanks to the seminal works by Cahill \& Pohl \cite{Cahill_1987, cahill1990} and Birge \& Nagel \cite{Birge1987}, the periodic heating concept was shaped into a suitable method for determining thermal conductivity in both bulk and thin films, currently known as $3\omega$ technique. The name ``3-omega'' derives from the fact that the third harmonic voltage is measured to extract the sample's thermal conductivity, and hence it can be classified as a ``frequency-domain'' technique.

The $3\omega$ method has been widely used for thermal conductivity experiments and adapted to investigate an extensive array of systems, which includes thin films \cite{Quintana2008, Park2014, Paterson2020}, bio-based polymers \cite{Boussatour2018}, biological tissues \cite{Madhvapathy2022, Ouyang2024}, nanowires \cite{Hasegawa2013}, nanotubes \cite{Lu2002, Kong2018} and liquids \cite{Lee2009, Schiffres2011}. In the framework of solid materials, the accurate determination of thermal conductivity can be crucial to assess the performance for thermoelectric generators, energy storage (e.g. batteries, thermal coatings), and microelectronics \cite{zheng2021advances_energy, ohtomo2022electrochemical, moore2014_thermalmanag}.
In some cases, elevated thermal conductivity of the components is needed (e.g., for an efficient dissipation of heat). In contrast, in other cases, a low conductivity is desired (e.g., for thermoelectricity). 

In this context, the 3$\omega$ method finds a proper and convenient application for directly obtaining the thermal conductivity of bulk materials, as it does not require prior knowledge of other optical and/or thermal parameters like in the case of techniques such as Laser Flash Analysis or Opto-Thermal Raman spectroscopy \cite{Parker1961, An2023}, which also require expensive instrumentation. 
Unlike steady-state methods, $3\omega$ is almost insensitive to radiative losses \cite{cahill1990}, therefore, it can be employed to explore the thermal conductivity of solids in a wide temperature range, typically from 50~K up to temperatures as high\cite{kawabata2024sapphire} as 850~K. Obtaining reliable data that do not need radiative correction, as in the case of the stationary methods, is a valuable advantage when the accurate estimation of thermal conductivity determines the suitability of a material for a specific application. 

The principles of the technique are available in several references, and they have been recently reviewed\cite{Bhardwaj2022}. However, from a practical point of view, many technical details fundamental to a prompt and successful methodology implementation are scattered among the references and often not even reported. This work aims to provide an in-depth discussion of the most crucial technical aspects underlying the implementation of this technique. We will first revise the essential concepts of the standard $3\omega$ method in Section \ref{sec:3omega}. In Section \ref{sec:guide}, we scrutinize the four main steps to get ready with the method. To support the design phase, we provide a Python-based graphical user interface\cite{pythonInterface} that allows for a quantitative estimation of the investigated temperature profiles based on the geometrical parameters (width/length) of the deposited heater/thermometer metal line prior to an experiment (Section \ref{sec:tool}). In Section \ref{sec:electronics} we describe an electronic configuration based on two lock-in amplifiers, often available in condensed matter laboratories. In Section \ref{sec:results}, we exemplify a thermal conductivity measurement performed on SrTiO$_3$ commercial readily-available insulating test samples, obtained using the proposed circuitry arrangement. Finally, we revisit the uncertainty expressions (Section \ref{sec:error}) to suggest which parameters should be optimized to ensure high-resolution results.

It is important to remark that the steps suggested in this work regarding the preparation of a sample for a thermal conductivity experiment based on the 3$\omega$-technique are straightforward only in the case of an insulating sample. The case of electrically conducting samples is briefly discussed. 

\section{The 3$\omega$-technique}
\label{sec:3omega}
In this section, we revise the main principles and fundamental formulas of the 3$\omega$ technique when applied to insulating, bulk samples. In Section \ref{sec:fundamental}, we report the general formulas, while in Section \ref{sec:linear} we discuss the linear approximation that is generally used in analysis of the experimental data to extract $\kappa$ from the frequency sweeps.

\subsection{Fundamental concepts}
\label{sec:fundamental}
\begin{figure}[ht]
\centering
\includegraphics[width=0.45\textwidth]{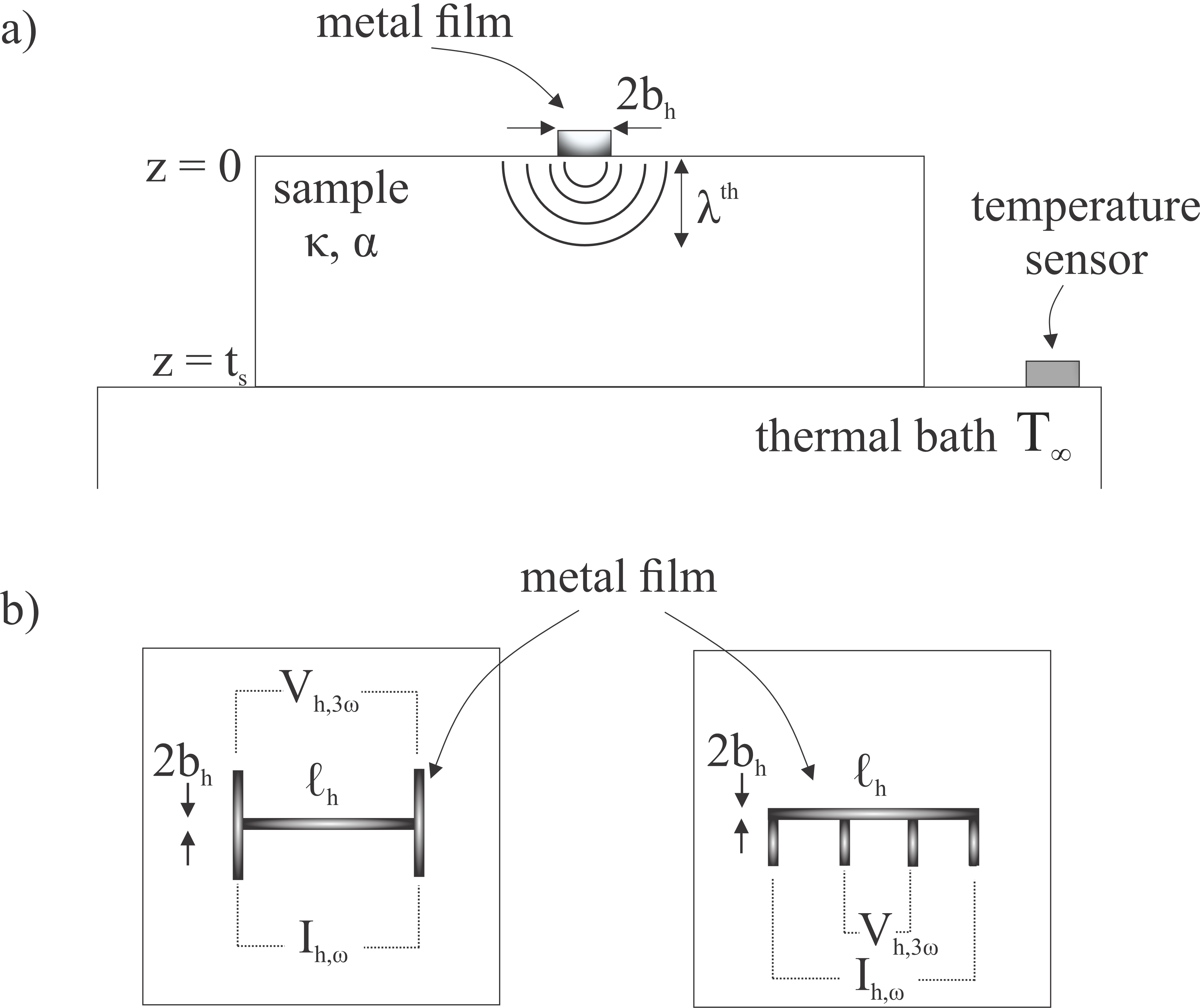}
\caption{a) Standard geometry for a 3$\omega$ experiment. The grayish thin film on top acts as heater and thermometer, and normally has four contact pads serving as current input ($I_h)$ and voltage measurement ($V_{h, 1\omega}$). Sample's surface is taken as $z=0$ plane. Thermal conductivity can be obtained from detecting the small third harmonic voltage generated from an alternating current $I_h = I_0\cos(\omega t)$ fed to the metal film. Isothermal condition at $z=t_s$ is assumed, given the thermal bath at temperature $T_{\infty}$ and intimate thermal contact between sample and bath. 
b) Top view of the metal lines deposited on top of the sample. Two geometries are usually chosen, as discussed by Cahill \cite{cahill1990}. $\ell_h$ represents the distance between current contact pads, as the entire H/T length contributes to the total dissipated power.
}
\label{fig:Fig1}
\end{figure}

 In the standard $3\omega$ technique, a metal film deposited over a sample acts simultaneously as a heater and resistance-thermometer (H/T), as exemplified in Figure \ref{fig:Fig1}a). If an alternating current flows through it, the sample is periodically heated due to the Joule effect. The sample is kept in vacuum, thus, no power is lost through convection. The resistance of the metallic line responds to this heating with frequency $2\omega$. Thus, a third harmonic in the voltage signal proportional to $R(2\omega) I(\omega)$ is measured. The dependence of the amplitude and phase of the oscillations with respect to the frequency can be related to the sample's thermal conductivity. Subscript ``$h$'' indicates a property of the H/T.

Let us consider an alternating current 
\begin{equation}
\centering
\label{eq-1}
I_h(t) = I_0\cos(\omega t)
\end{equation}
with frequency $\omega$ flowing through a metallic thermal resistor, a material whose resistance has a linear temperature dependence 

\begin{equation}
\centering
\label{eq0}
R_h = R_0(1+\beta_h \Delta T),
\end{equation}
where $R_0$ is the reference resistance and $\beta_h$ is the thermal coefficient of resistance (TCR). The dissipated power along the resistor depends on both $1\omega$ and $3\omega$ voltage harmonics, the former being a constant and the latter an oscillating component \cite{koninck2008, Pradipta2016}. The metallic line has a full-width $2b_h$, length $\ell_h$ and is assumed to have negligible thickness $t_h$, where $t_h \ll t_s$ --- see Figure \ref{fig:Fig1}b). 

Measurements of the voltage at frequency $3 \omega$ allow us to determine in-phase (real) and out-of-phase (imaginary) components of temperature oscillations, which relate to the sample's thermal conductivity $\kappa$. The amplitude of the $3\omega$ voltage component is expressed as

\begin{equation}
\centering
\label{eq2}
   V_{3\omega} = \frac{1}{2}\beta_h V_{1\omega} T_{2\omega}  
\end{equation}
where $V_{1\omega} = R_0I_0$ is the first harmonic voltage and $\beta_h$ is given by

\begin{equation}
\centering
\label{eq2.5}
   \beta_h(T) = \left[\frac{1}{R_h(T)} \frac{dR_h}{dT}\right]\bigg|_{T=T_{\infty}}
\end{equation}
and $P_{\text{DC}} = V_{1\omega} I_0 = V_{1\omega}^2/R_h$ is the DC dissipated power. We highlight that some references\cite{DKim2004, Hahtela2015} refer to the TCR as only the sensitivity $\frac{dR_h}{dT}$, which can cause inconsistencies when comparing data from multiple sources. We call attention to the fact that this derivative in Equation (\ref{eq2.5}) is not constant over all temperatures: for most pure metals (Pt, Au, Ag, etc) that are often employed, the resistances curves flatten as temperatures approach $\sim 50$ K, leading to the vanishing of the TCR coefficient, and thus to the impracticability of the $3\omega$ method below that threshold. New alloys or alternative materials, including ZrN$_x$ and Si:Nb \cite{Yotsuya1987, Vecchio1995}, have been proposed as candidates to pursue low temperature $3\omega$ thermal conductivity measurements.

It can be demonstrated \cite{Banerjee1999, Hanninen2013, koninck2008} that the temperature on the surface of the sample, $T_{2\omega}$, due to the finite dimensions of the heater strip, is given by
 
\begin{eqnarray}
\centering
\label{eq1}
   T_{2\omega} && = \frac{P_{\text{DC}}}{ \pi \ell_h \kappa} \int_{0}^{\infty}\frac{\sin^2(\eta b_h) }{(\eta b_h)^2\sqrt{\eta^2 + |\mathbf{q}|^2}}d\eta  \\&& = \Re(T_{2\omega}) + i \Im(T_{2\omega})
\end{eqnarray}
where the thermal wave-vector is $|\mathbf{q}|^2 = 2\omega i / \alpha$, $\alpha = \kappa/C_p$ the thermal diffusivity ($C_p$ is the heat capacity) and $\eta$ is an integration variable. The last equality comes from the fact that $\mathbf{q}$ is a complex number; thus, $T_{2\omega}$ has both real and imaginary parts. We can think of $T_{2\omega}$ as a temperature field \cite{Ordonez-Miranda2023}: that expression relates the temperature at any penetration depth,

\begin{equation}
    \label{eqlambda}
    \lambda^{\text{th}} \doteq \frac{1}{|\mathbf{q}|} = \sqrt{\frac{\alpha(T)}{2\omega}},
\end{equation}
given the H/T frequency $\omega$ through the thermal wave-vector $q$. The expression was also generalized for anisotropic samples with independent heaters and sensors \cite{Ordonez-Miranda2023}. $T_{2\omega}$ is sometimes\cite{Jaber2018, Xing_2014} referred to as $\Delta T (2\omega) = T_{2\omega}(z=\lambda^{\text{th}}) - T_{\infty}$, representing the change of the temperature amplitude at a depth $z = \lambda^{\text{th}}$ with respect to the environment temperature $T_{\infty}$. Equation (\ref{eq1}) implicitly assumes semi-infinite boundary condition (semi-infinite sample thickness).

\begin{figure}[htbp]
\centering
\includegraphics[width=0.45\textwidth]{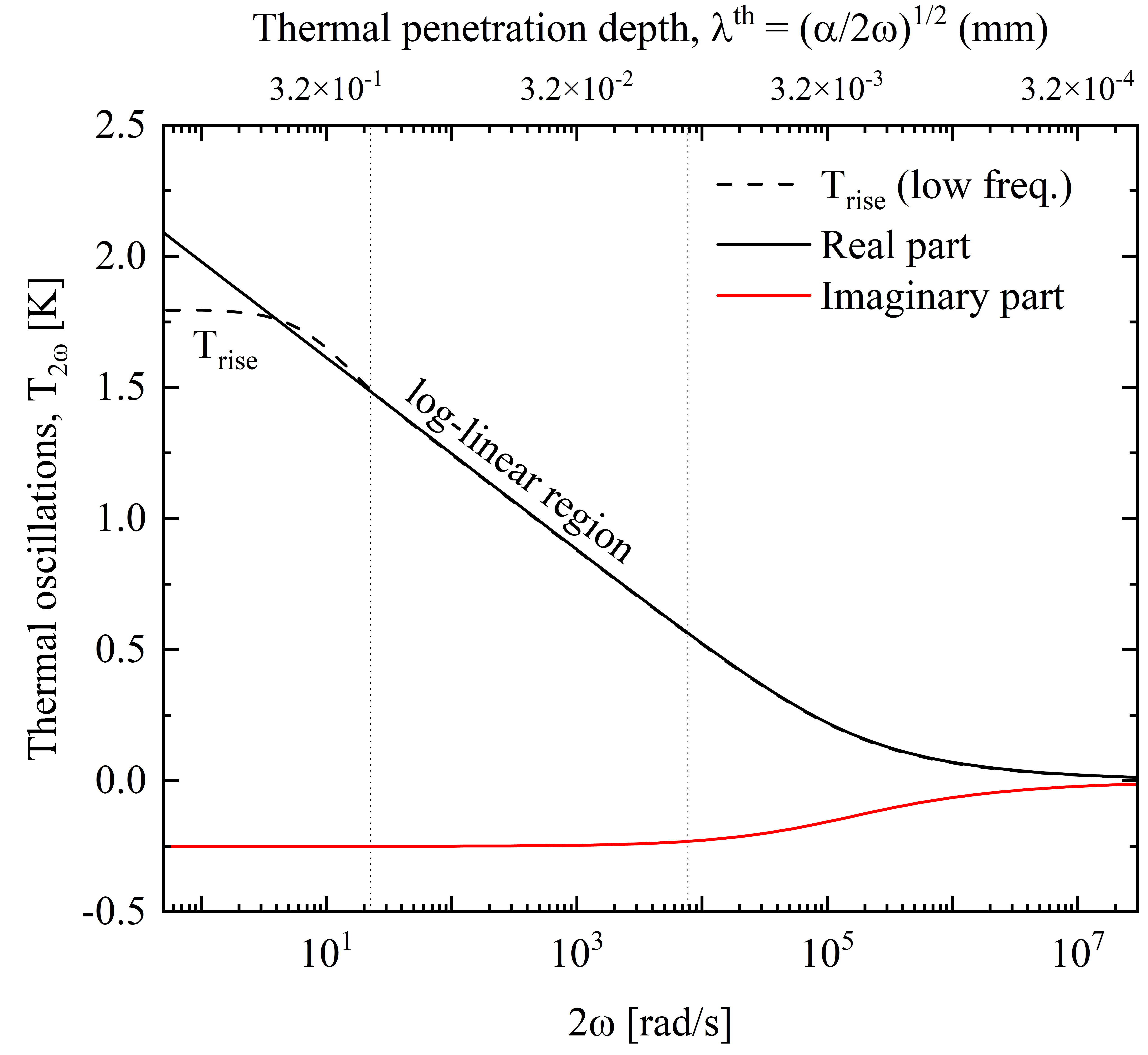}
\caption{\label{fig:linear} Real and imaginary parts numerically solved for Equation (\ref{eq2.5}) with the parameters as specified in the text. The dashed vertical lines indicate the log-linear region. $T_{\text{rise}}$ is calculated as discussed in the text.}
\end{figure}

Let us analyze the meaning and implications of Equation (\ref{eq1}), which does not have an analytical solution. Numerical integration allows for simulating the profile of temperature oscillation amplitudes as a function of the excitation frequency. An alternative formulation of the temperature field in Equation (\ref{eq1}) in terms of G-Meijer functions was presented by Duquesne and co-workers \cite{Duquesne2010}. For illustration, Figure \ref{fig:linear} shows a simulation of the real and imaginary parts of temperature oscillations as a function of twice the input frequency, calculated by using Equation (\ref{eq1}). To compute the integral, $P_{\text{DC}}$ was set to $1$~W, $\ell_h$ to $1$~m, $\kappa$ to $1$~W m$^{-1}$K$^{-1}$, heater half-width $b_h$ was set as 5 $\mu$m and $\alpha$ to $1$ mm$^2$s$^{-1}$. The sample thickness was set to $t_s = 0.5$ mm. In practice, the integral is calculated over the interval $10^{-10} < \eta < 10^{7}$. Increasing this range just changed the amplitude by less than 1\%; therefore, we have restricted the simulation to those limits. The simulation was performed using a Python-based software that we will discuss in Subsection \ref{sec:tool}.

As shown in Figure \ref{fig:linear}, for $\lambda^{\text{th}} \gg b_h$, the real part of thermal oscillations decays linearly in the semi-log plot, hence the name ``linear'' regime. This regime also has a slowly varying and negative imaginary part. At small penetration depth (high frequencies), an asymptotic behavior occurs ($T_{2\omega} \rightarrow 0$), where the in- and out-of-phase components are equal but with opposite signs, representing the lag between current and temperature.

If the sample is considered semi-infinite ($t_s \rightarrow \infty$, or equivalently, $\lambda^{\text{th}} \ll t_s$) 
, the thermal oscillations should diverge for $2\omega \rightarrow 0$. However, as the temperature oscillations have to eventually decay far away from the heater $\lambda^{\text{th}} \gg b_h$, we might impose $T_{2\omega} \rightarrow T_{\text{rise}} = P_{\text{DC}}/( \pi \ell_h \kappa)\left( \log(t_s/b_h) + 1.0484 \right)$ for very low excitation frequencies and finite sample thickness, following the discussion of Jaber \& Chapuis \cite{Jaber2018}. The latter is represented by the dashed line in Figure \ref{fig:linear} at low frequencies.

\subsection{Linear approximation}
\label{sec:linear}
The previous section shows how the frequency response of the real and imaginary parts of $T_{2\omega}$ can be numerically solved for Equation (\ref{eq1}). Figure \ref{fig:linear} represents the outcome of an ideal experiment, carried out with the parameters specified above. In a real experiment, the slope of the log-linear region of the real part is typically used to obtain $\kappa$. Therefore, in principle, it is preferable to choose dimensions that maximize the linear regime range \cite{Luo2006}. Let us revise the conditions to experimentally obtain a sufficiently extended linear range, given a certain sample geometry.

Based on Equation (\ref{eq1}), if the width of the heater is small enough to satisfy the condition $qb_h << 1$, one can approximate the real part of $T_{2\omega}$ as 

\begin{equation}
\centering
\label{eq2.7}
   T_{2\omega} = -\frac{P_{\text{DC}}}{ 2 \pi \ell_h \kappa} \left[ \log(2\omega) + \frac{1}{2}\log(\frac{b_h^2}{\alpha}) - 2\xi \right]
\end{equation}
which has linear behavior in log-space, where $\xi= 3/2-\gamma$ ($\gamma$ is the Euler-Mascheroni constant). In this case, the thermal conductivity is given by\cite{Kaul2006} the expression

\begin{equation}
\label{eq3}
    \kappa = \frac{I_h^3R_h\beta_h}{4\pi\ell_h}\left(-\diff{V_{3\omega}}{\log(2\omega)}\right)^{-1}
\end{equation}



The thermal penetration depth (plotted in the upper scale of Figure \ref{fig:linear}) indicates how deep thermal oscillations penetrate the specimen before their amplitude gets strongly damped. Thus, the linear approximation is reasonable when the thermal penetration depth is sufficiently large compared to the H/T width (so that the 1D heat flow is a reasonable assumption) and small compared to the sample thickness (preventing back-reflection of thermal waves). Such a condition is often summarized as 

\begin{equation}
\centering
\label{eq4}
   b_h \ll \lambda^{\text{th}} \ll t_s.
\end{equation}

A more precise inequality might be reformulated as $n\times b_h < \lambda^{\text{th}} < t_s/n$, where $n$ is a number often found to lie between integers \cite{Gesele1997, koninck2008} $n = 1$ and $n=5$. The latter constraint was shown to bring a maximum error of 1\%, but may become impracticable due to its very restrictive geometry requirement. The actual $n$ must be evaluated for a given experiment, and it seems to be sample-dependent and/or sensitive to the heater-sample adhesion and sample-platform thermal contact. In most experiments, we have used $n=1$ and found reasonable results. 

\section{Methods: getting ready with 3$\omega$ technique}
\label{sec:guide}
\begin{figure}[htbp]
\centering
\includegraphics[width=0.45\textwidth]{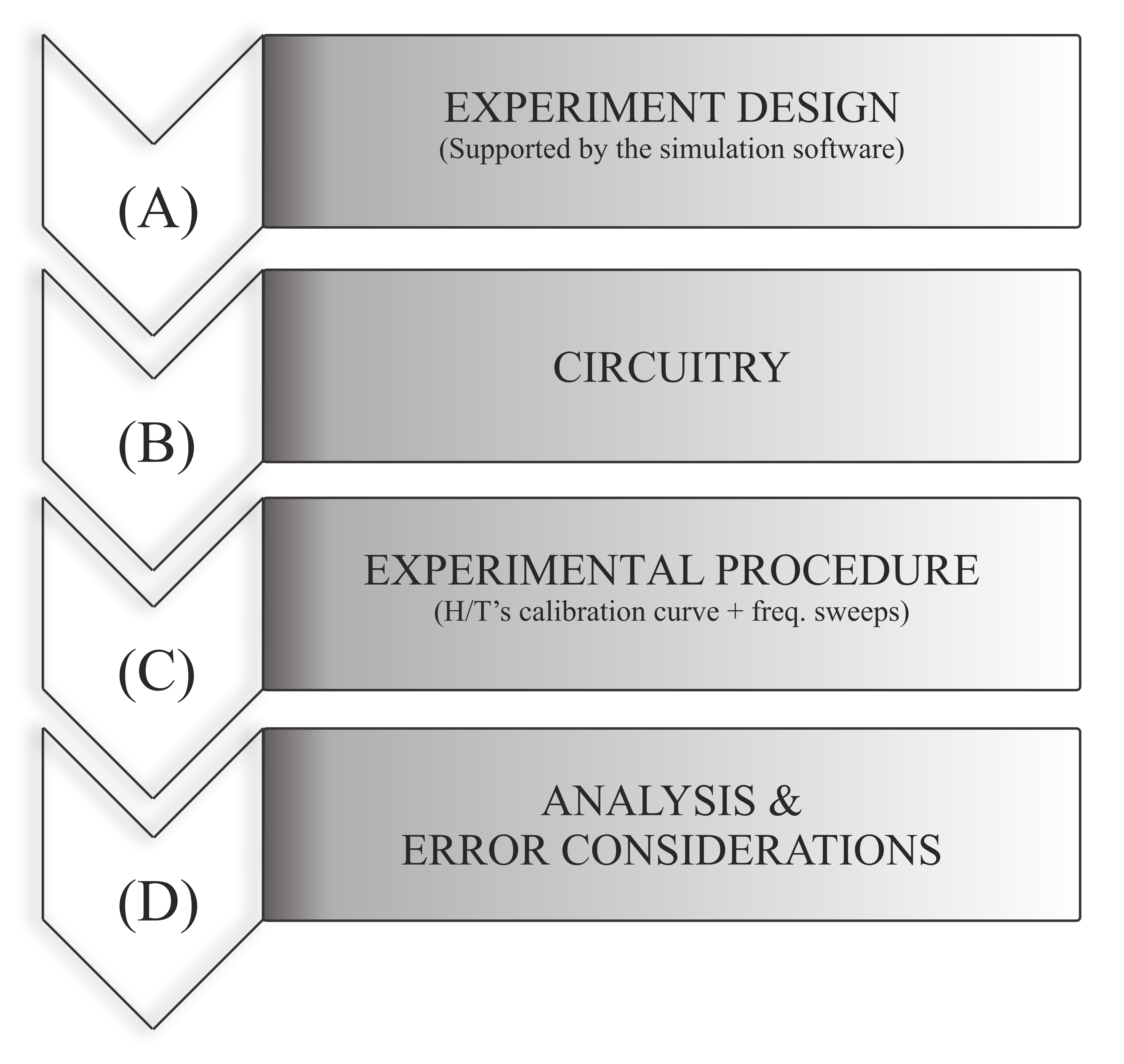}
\caption{\label{fig:steps} Essential steps for an optimized implementation of the 3$\omega$ technique on a bulk sample.}
\end{figure}

This section aims to provide practical guidance on the most relevant steps in the method's implementation, which are summarized in Figure \ref{fig:steps}. In Subsection \ref{sec:tool}, we discuss how to define the appropriate geometric parameters for the H/T transducer, before any experiment, when a sample of a specific geometry is selected. We propose using a Python-based interface\cite{pythonInterface} to perform a quick simulation of a possible 3$\omega$ response based on the choice of geometric parameters. The code is available for download in 
the repository of Ref.\cite{pythonInterface}, along with its documentation and an executable installer. We illustrate how to use this tool on selected standard crystals (ex. SrTiO$_3$) and demonstrate how an experiment can be planned using the provided interface. In Subsection \ref{sec:electronics}, we describe an electronic setup based on two lock-in amplifiers, providing useful quantitative references for the instrumental parameters. In Subsection \ref{sec:results}, we report on the essential steps of the measurement execution. Finally, in Subsection \ref{sec:error}, we analyze the main sources of errors, which point to the critical parameters that must be optimized to guarantee a measurement as accurate as possible.

\subsection{Experiment design: A Python-based tool for parameter estimation}
\label{sec:tool}
A critical step in planning a 3$\omega$ experiment is designing the heater/thermometer's geometry, which determines the extension of the linear region and, thus, impacts the measurement accuracy when the linear approximation is used. The heater/thermometer line has a typical geometry as reported in Figure \ref{fig:Fig1}(b), which contains two contacts for providing the current and two contacts for reading the voltage drop across the H/T. From a practical standpoint, the choice of the heater width is crucial for the experiment resolution. 

The input values of the Python-based software are: sample's thermal conductivity and thermal diffusivity, H/T line half-width, length and dissipated power $P_{\text{DC}}$, in addition to the minimum and maximum frequencies $2\omega$ for the simulations. Of course, in an experiment with non-characterized materials, estimates of their thermal parameters should be used to get started (see also discussion below). The outputs are the real and imaginary parts of $T_{2\omega}$, numerically solved from Equation (\ref{eq1}).  

\begin{figure}[htbp]
\centering
\includegraphics[width=0.45\textwidth]{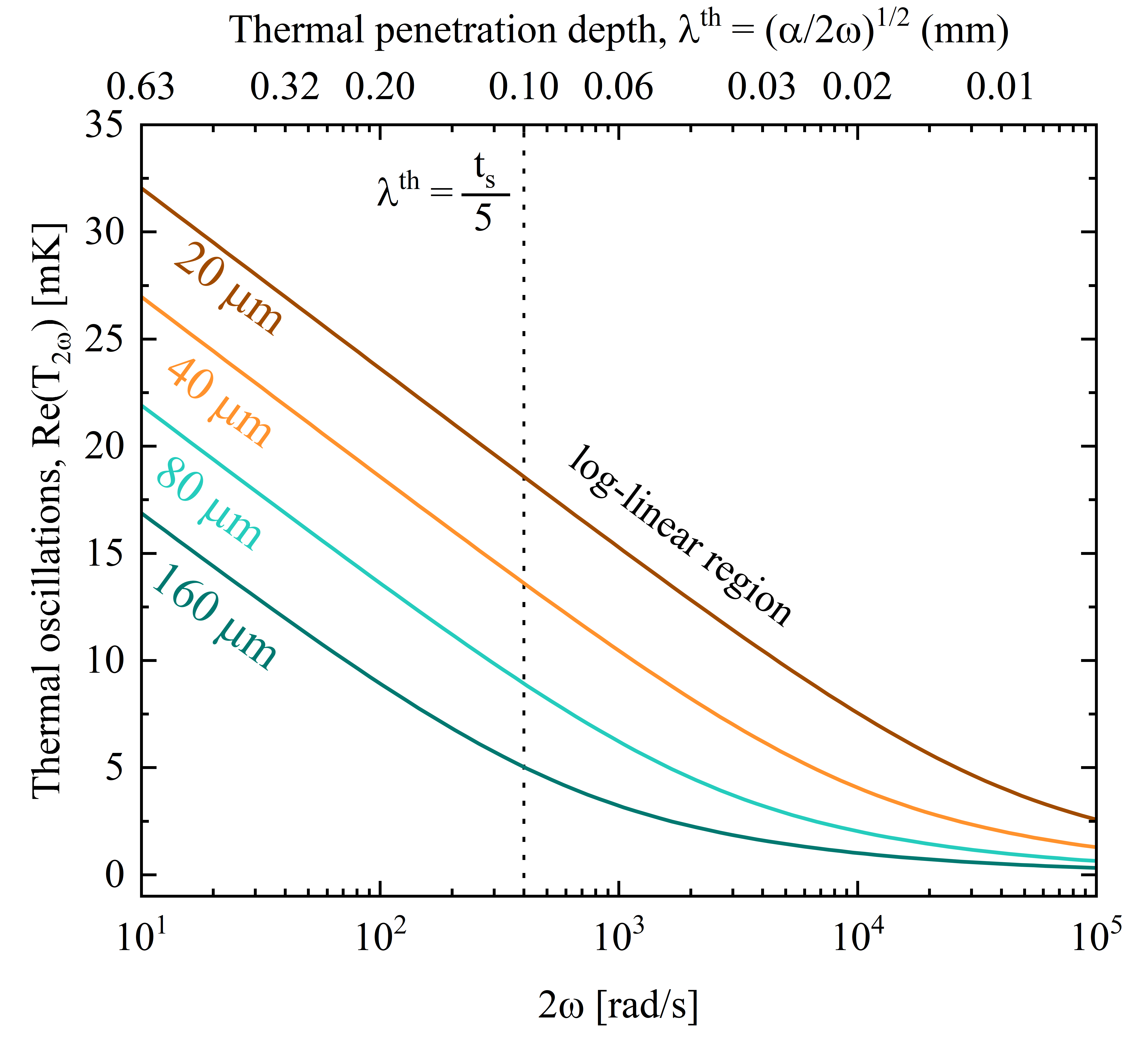}
\caption{\label{fig:various2b} Simulation of the real part numerically solved for Equation (\ref{eq1}) (as in Figure \ref{fig:linear}), in the case of a SrTiO$_3$ crystal for different line half-widths ($b_h =$ 20 $\mu$m, 40 $\mu$m, 80 $\mu$m and 160 $\mu$m). The vertical dashed line indicates where the thermal penetration depth is equal to one-fifth of the sample thickness, if the sample's thickness was $t_{\text{s}} = 0.5$ mm, following the discussion of constrains for the linear regime, Equation (\ref{eq4}). Note that the simulation implicitly assumes semi-infinite sample thickness. See the software manual for details on different boundary conditions that can be considered in the simulations.} 
\end{figure}

Let us analyze the case of a SrTiO$_3$ insulating sample. For this compound, we can input data available from literature at room temperature \cite{martelli_prl}: $\kappa_{\text{STO}} \sim $10 W m$^{-1}$K$^{-1}$ and $\alpha_{\text{STO}} =$ 4 mm$^{2}$ s$^{-1}$. In principle, two parameters regarding the H/T can be chosen: its half-width $b_h$ and length $\ell_h$. As $\ell_h$ will only affect the magnitude of the third harmonic signal, because of Equation (\ref{eq1}), we have fixed $P_{\text{DC}}/\ell_h = 0.23$ W/m. Thus, we can select different values for the line half-width and simulate the $T_{2\omega}$ response as a function of the excitation frequency. Figure \ref{fig:various2b} displays the output for line half-widths of 20, 40, 80, and 160 $\mu$m. We can inspect the range of extension of the linear regime for each width. In the specific case we chose for the simulation (SrTiO$_3$), we can see that, in order to have an extended linear region where the thermal penetration depth is still confined inside the sample, a line thinner than $\sim$ 40 $\mu$m should be selected. 

Thinner widths allow for an extended linear region towards higher frequencies, which in principle suggests that the thinner the line, the more accurate the determination of the slope becomes. Extending the linear range towards lower frequencies requires thicker samples, once the H/T geometry has been defined. While this holds in theory, in practice, an experimentalist making this decision must consider that reducing the line width further might imply using lithography rather than sputtering for the H/T fabrication. Sometimes, access to lithography can be more difficult and/or the sample could be sensitive to the chemicals used in the lithography process. Furthermore, the compliance of the available electronic instrumentation will limit the upper value of frequency that can be supplied. The lower limit in frequency has to be chosen considering that the thermal penetration depth should be less than $t_{\text{s}}/n$. In summary, our software\cite{pythonInterface} supports the user's decision-making process, providing a simulated plot as in Figure \ref{fig:linear} for given specifications of a sample. 

It is possible that a specific material of interest does not have previous reports on the thermal parameters: this is often the case when the specimen is a prototype. In this circumstance, it is possible to choose a material expected to be structurally and chemically similar to the one under investigation. The analysis will not be exact, but the estimation of the line geometry may be good enough. As a last resort, if this estimation is not viable, one can perform a frequency sweep over all the range available, preferably on thick samples and thinner H/T lines, to identify the actual frequency boundaries of the linear range. It is worth noticing that, as the thermal diffusivity of any sample changes as a function of temperature, the minimum and maximum frequencies $2\omega$ of the linear range also change, as can be seen by combining Equations (\ref{eqlambda}) and (\ref{eq4}).


\subsection{Circuitry: the two lock-in configuration}
\label{sec:electronics}

This section is devoted to the application of the methodology to a circuitry based on the use of two Lock-in Amplifiers. It is worth noting that other sets of electronic instruments can serve the purpose equally well\cite{cahill1990, Park2014, Boussatour2018, Chernodoubov2019}, when the circuit is properly adapted. Here, we chose to work with two Lock-in Amplifiers, being instruments typically available in most laboratories. 

The temperature oscillation $T_{2\omega}$ can be experimentally obtained by measuring $V_{3\omega}$, according to Equation (\ref{eq2}). 
Since the thermal conductivity is determined in terms of the slope of $V_{3\omega}$, according to Equation (\ref{eq3}), it suffices to investigate the real part of the third harmonic voltage. 
Notice that, since $\beta_h \sim 10^{-3}$ K$^{-1}$ for most of the pure metals often employed, it is not uncommon to have $V_{3\omega}/V_{1\omega} \sim 10^{-5}$. Although lock-in amplifiers can recover any harmonic signal buried in a given input voltage, suppressing the 1st harmonic is highly recommended for increasing the measurement's resolution. 

\begin{figure}[htbp]
\centering
\includegraphics[width=0.52\textwidth]{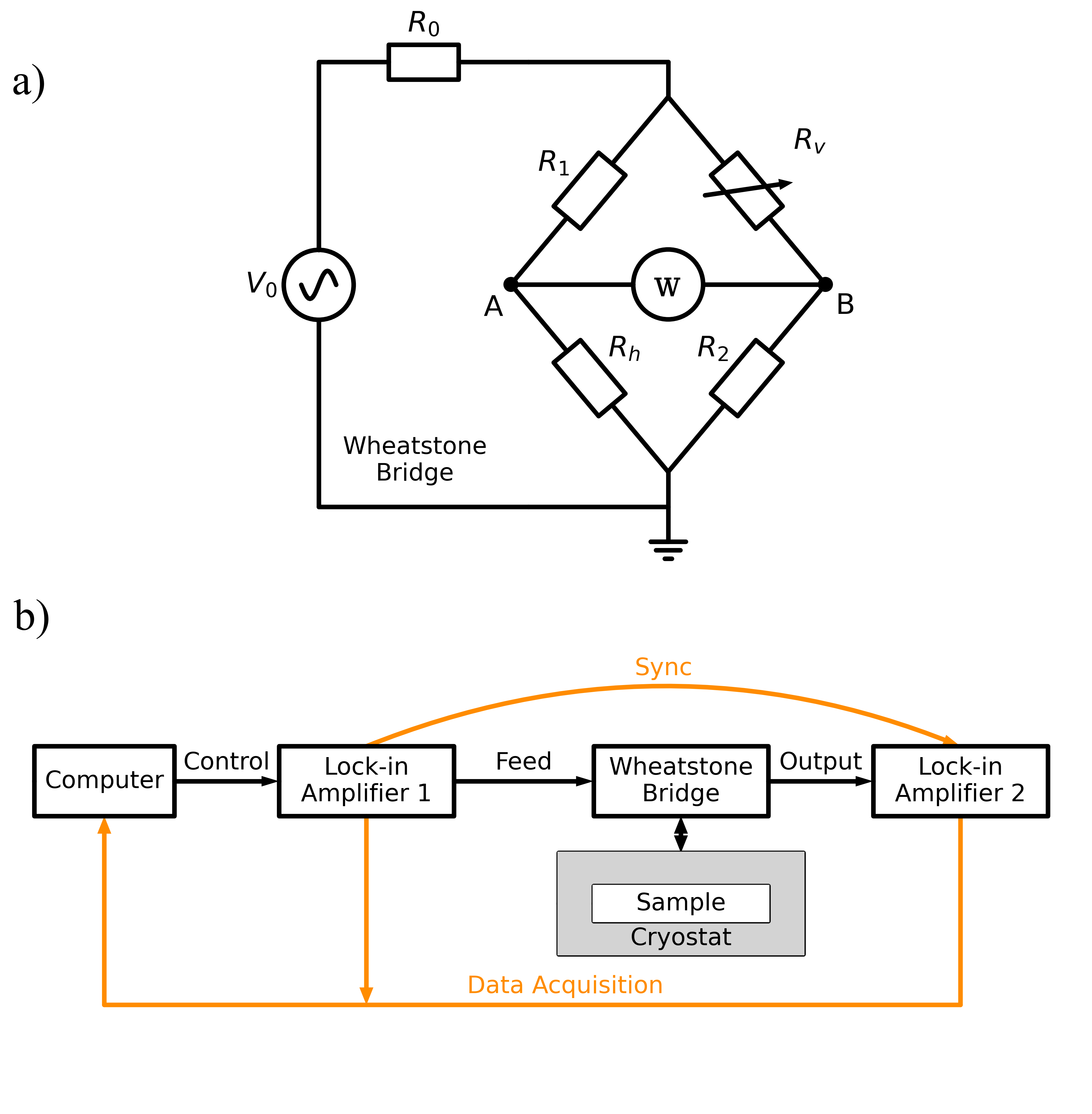}
\caption{a) Flowchart of the $3\omega$ electronics, and b) depicting of the electronic circuit based on a WB. The total signal is $W = W_{1\omega} + W_{3\omega}$. The Lock-in Amplifier 1 is responsible for voltage input $V_0$ (signal generator) and first harmonic measurement, while the Lock-in Amplifier 2 detects the third harmonic $W_{3\omega}$. Both Lock-in amplifiers are synchronized for a common frequency $\omega$ reference. See Table \ref{tab1} and text for details.}
\label{fig:flowchart}
\end{figure}

The suppression of the first harmonic voltage can be obtained by introducing a Wheatstone Bridge (WB), as seen in Figure \ref{fig:flowchart}a). 
For a given $R_h$, and by fixing the resistances $R_1$, $R_2$, the first harmonic signal is suppressed by varying the resistance $R_v$. $R_1$ is the value of the in-series resistor. If the heater is the only element in the bridge that generates a third harmonic voltage drop, then balancing the bridge will attenuate the first harmonic signal without affecting the $3\omega$ voltage component \cite{Yamane2002, Hu2006, koninck2008}. Close to the balancing condition,

\begin{equation}
    \label{eq:tuneWB}
    R_v = \frac{R_1R_2}{R_h}
\end{equation}
is a useful initial guess for the resistance required for first harmonic voltage suppression. As $R_h$ varies with temperature, the variable resistance must be adequately adjusted for temperature-dependent experiments. This variable resistor could be a commercial potentiometer, but for long-term stability and precision, we have employed a digital resistance decade PRS-330, that spans a wide range covering a few m$\Omega$ to several M$\Omega$ (check Table \ref{tab1}). If a programmable decade is not available, a manual/analogic decade can be employed instead. 

As a consequence, it can be shown\cite{koninck2008} that the suppression of the first harmonic voltage leads to a rescaled third harmonic signal 

\begin{equation}
    \label{eq:W}
W_{3\omega} = \frac{R_1}{R_1+R_h}V_{h, 3\omega}
\end{equation}
by treating the bridge as a voltage divider. Based on our results, an attenuation down to $V_{3\omega}/V_{1\omega} \sim 10^{-3}$ is sufficient for clear third harmonic readings.

Given that the third harmonic voltage depends on $I_h^3$ according to Equation (\ref{eq3}), it is desirable that most of the current split between the arms of the WB flows through the H/T. Denoting $I = I_L +I_R$ the magnitudes of current flowing through the Left and Right arms of the WB in Figure \ref{fig:flowchart}a), we can derive that $I_L/I_R = (R_2+R_v)/(R_1+R_h)$. Therefore, one can maximize the current flowing through the H/T by properly selecting the fixed resistances $R_1$ and $R_2$. Considering the example above, if $R_h = 37\; \Omega$, then selecting $R_1=50\; \Omega$ and $R_2=2\; k\Omega$, by assuming that $R_v = 2.7 \;k\Omega$ from Equation (\ref{eq:tuneWB}), we would have $I_L/I \sim $ 98\%.

The complete diagram of the used setup for $3\omega$ measurements is presented in Figure \ref{fig:flowchart}b). An oscillating voltage $V_0 \cos(\omega t)$ provided by a Lock-in amplifier-\#1 [in Figure \ref{fig:flowchart}b)], feeds the WB. This lock-in also measures the fundamental voltage across the H/T, which is an integral part of the WB, Figure \ref{fig:flowchart}a). A second Lock-in Amplifier-\#2 [Figure \ref{fig:flowchart}b)] records the voltage between the WB arms, which is the rescaled third harmonic voltage $W_{3\omega}$ according to Equation (\ref{eq:W}). All the data is stored in a computer, which is also employed to send commands to all the equipment. The complete list of instruments used in the setup of the technique is presented in Table \ref{tab1}. It is worth pointing out that the approach we discussed in this work to obtain the suppression of the first-harmonic voltage is one of the possible choices and can be replaced by other alternative electronic configurations\cite{Kaul2006, Park2014, Boussatour2018} aiming at optimizing the $V_{3\omega}/V_{1\omega}$ ratio.

\begin{table*}[t]
\begin{ruledtabular}
{\renewcommand{\arraystretch}{5}}
\caption{List of instruments used in the $3\omega$ technique implementation.}\label{tab1}%
\begin{tabular}{@{}ccc@{}}

Company & Model  & Purpose \\
\hline
Signal Recovery    & 7270 Dual Phase  Lock-in Amplifier    & \makecell[c]{Signal reference and\\ first harmonic measurement.}   \\
Signal Recovery    & 7270 Dual Phase  Lock-in Amplifier   & \makecell[c]{Signal generator and\\ third harmonic measurement.}  \\
Lakeshore    & \makecell{331 Cryogenic \\ Temperature Controller}   & Temperature control.  \\
Lakeshore    & 372 AC Resistance Bridge   & Measuring the H/T resistance, $R_h$.  \\
IET Labs &  \makecell[c]{PRS-330 Programmable \\ Resistance Box} & Balancing the WB, $R_v$.\\

Advanced Research Systems   & DE-202NI Cryocooler   & \makecell[c]{Eliminating convection effects\\ and cooling power.} \\

\end{tabular}

\end{ruledtabular}
\end{table*}



To verify the working principle of the technique, we selected bulk, insulating SrTiO$_3$ and LiNbO$_3$ ([100]-oriented; thickness $t_s = 0.5$ mm, commercially available) as test-bed compounds. Metallic lines were fabricated by shadow-masking the patterns through RF-sputtering of Pt (LiNbO$_3$) and lithography of Au (SrTiO$_3$). It is important to advise that rather electrically conducting and/or easily degradable samples must be coated with a thin layer of an insulating material before-hand, in order to proceed with the lithography process for the H/T fabrication. Many materials are reported as suitable choices for capping layers, as SiO$_2$ \cite{Kaul2006}, Al$_2$O$_3$ \cite{Cui2017} and Si$_3$N$_4$ \cite{Qiu2021}. Such a layer is needed to ensure that the
metallization for the heater/thermometer test pattern is adequately insulated from the sample when the situations aforementioned happen \cite{Lee1997}.

The thickness of the H/T, $t_h$, may not influence on the thermal conductivity measurements (i.e. neglecting the contributions from the thermal mass of the H/T to the total heat capacity ``sample + H/T'') as long it is kept as thin as possible relatively to the sample's thickness. For the impact of a thick H/T in $3\omega$ thermal conductivity experiments, see the discussion elsewhere \cite{Dames2013, Jaber2018}. Considering the values reported\cite{cahill1990, Zhao2007} ranging from 20 nm to 300 nm, we have chosen $t_h\sim$100 nm. An intermediate layer of Cr or Ti (2 - 5 nm) might be useful for improved adhesion \cite{cahill1990, Zhao2007}. 

Platinum and gold were selected due to their relatively high TCR at room temperature (RT). Electrical contacts were prepared using 50 $\mu$m gold wires (Premion, purity 99.995\%) and Dupont silver paint. A Cernox CX-SD 1050 resistance sensor monitors the temperature of the platform, $T_{\infty}$. The Au-heater length $ \ell_h=3.70\pm0.01$~mm and full-width $2b_h=40\pm2$~$\mu$m were determined by analyzing high-resolution images with a Leica M205C stereoscope. We stress that the parameters were selected to meet the requirements found in the simulation. For LiNbO$_3$, only its RT thermal conductivity was measured (see Table \ref{tableResults}), so the following discussion is primarily based on the results from SrTiO$_3$.

In $3\omega$ experiments, the H/T power is adjusted such that the resulting temperature oscillation $T_{2\omega}$ is assumed to be a small fraction of the environment temperature, allowing $T_{\infty}$ to be regarded as representative of the measured thermal conductivity. While this approximation is generally valid, it must be noted that the oscillatory $2\omega$ heating component is inherently superimposed on a background DC power. While a minimum $p_0 > 800 \text{mW/m}$ is required to obtain measurable signals\cite{koninck2008}, the input power cannot be boundlessly increased, since otherwise this would induce significant self-heating of the H/T, leading to temperature drifts and consequent violation of the assumption above. 




\subsection{Experimental procedure}
\label{sec:results}
\begin{figure*}[ht]
\centering
\includegraphics[width=1.0\textwidth]{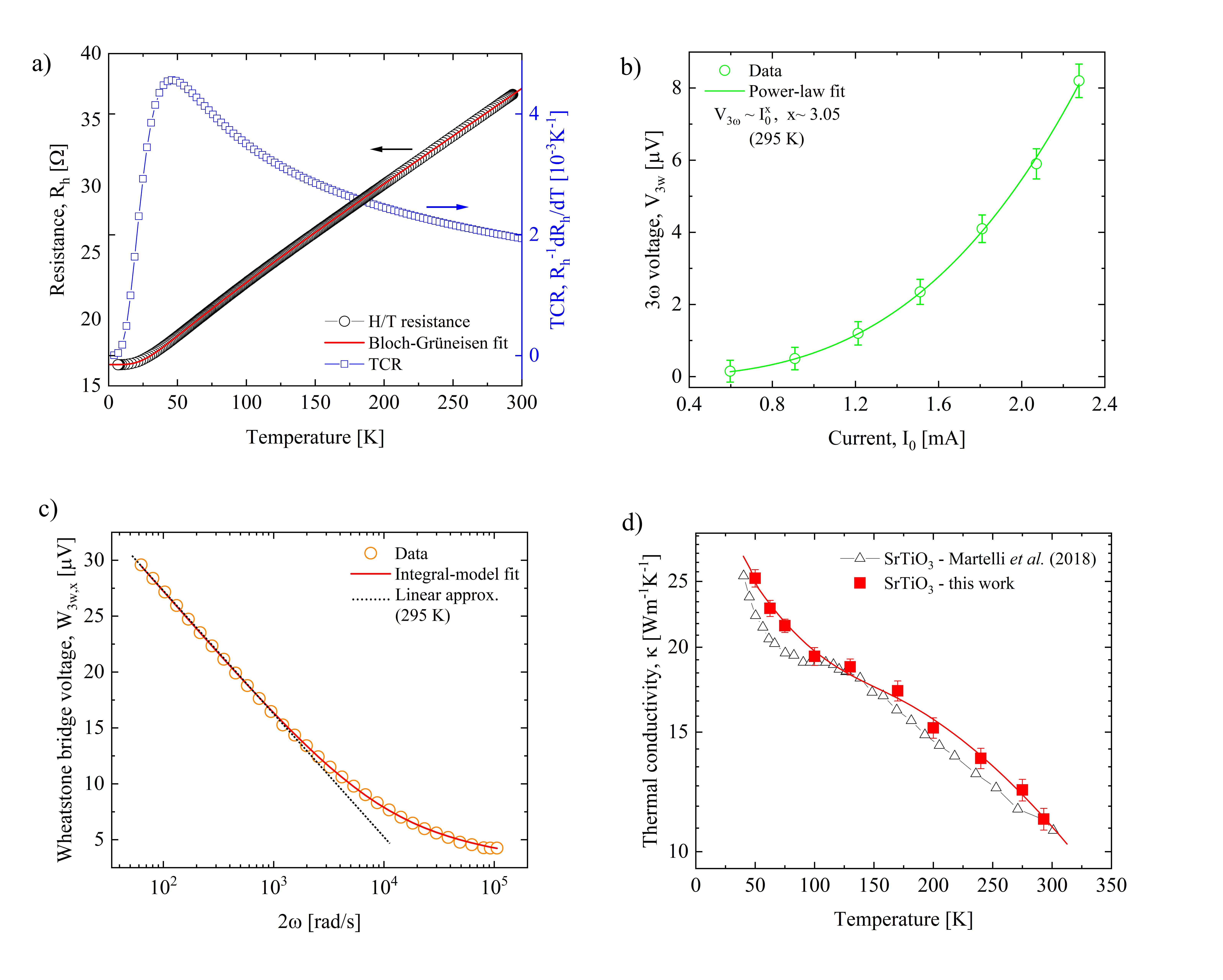} 
\caption{a) Measured variation of the H/T's resistance (Au) as a function of temperature, open circles, and its TCR (blue squares). Red solid line results from a fit of the Bloch-Grüneisen equation to the data. b) Evolution of $V_{3\omega}$ as a function of the current $I_0$ that crosses the heater/thermometer, demonstrating the cubic dependence expected from Equation (\ref{eq3}). c) Example of frequency sweep (open circles) and non-linear fitting  (solid line) at $T$ = 295 K. Uncertainties were found to be about 1\%, being not visible in the plot scale.  d) Thermal conductivity of SrTiO$_3$ as determined by the $3\omega$ method. The solid red line is a B-spline and serves as a guide for the eyes. Data from SrTiO$_3$ as measured via the standard thermal conductivity method --- open triangles --- are presented for comparison \cite{martelli_prl}.}
\label{fig:kappa}
\end{figure*}

\textit{Heater/Thermometer calibration}. The first measurement step is the calibration of the heater/thermometer deposited on top of the sample, to determine $R_h$ and $\beta_h$. Figure \ref{fig:kappa}a) exhibits the calibration curve of the Au heater/thermometer, obtained for our H/T deposited on SrTiO$_3$. An accurate description of the temperature dependence of the resistance is achieved by fitting the Bloch-Grüneisen (B-G) equation from semi-classical Boltzmann transport theory for metals \cite{Ziman1960, Bid2006, Putnam2003, Chen2009, Sawtelle2019},

\begin{equation}
\centering
\label{eqBG}
   R_h(T) =  R_{\text{const}} + K_{\text{ph-el}}\left(\frac{T}{\Theta_R}\right)^n\int_{0}^{\Theta_R/T} \frac{t^n}{(e^t-1)(1-e^{-t})}dt
\end{equation}
where $\Theta_R$ is the characteristic B-G temperature, typically matching the Debye temperature $\Theta_D$. For many transition metals, $n = 3$. $R_{\text{const}}$ is the low temperature saturation of the metal resistance and $K_{\text{ph-el}}$ is the electron-phonon coupling constant. The red line in Fig. \ref{fig:kappa} represents the fit obtained by Equation (\ref{eqBG}).

One can see that the heater/thermometer resistance starts to flatten below about $\sim 50$ K, or conversely, its TCR drops to zero --- blue squares, right-axis of Figure \ref{fig:kappa}a) --- evidencing the temperature at which the method becomes impracticable. The TCR of the resistance is computed by numerical differentiation of the analytical model. Therefore, for every single H/T it is recommended to fit the Bloch–Grüneisen formula containing three adjustable parameters to the measured $R_h(T)$, adopting the procedure taken in \cite{Putnam2003, Chen2009} for instance, with the objectives of \textit{i)} reducing the uncertainties in the TCR coefficient and \textit{ii)} getting to a more satisfying fit to the temperature-dependence of the H/T's resistance. Technically, measurements could be pushed down to even lower temperatures $\sim 20$ K by fitting the calibration curve within finite temperature intervals \cite{Dames2013}, but doing so may substantially increase uncertainties for $\beta_h$.

Notably, Figure \ref{fig:kappa}b) confirms the expected cubic power-law dependence of the 3rd harmonic voltage as a function of the current --- or interchangeably as a function of the first harmonic voltage, since $V_{h,1\omega} \approx R_hI_0$ --- at 295 K, as predicted by Equation (\ref{eq3}). 

\textit{3$\omega$ frequency sweeps}. Figure \ref{fig:kappa}c) shows a typical (real part) $W_{3\omega}$ frequency sweep at a fixed temperature, in this case $T$ = 295 K. We have found no significant difference between the non-linear fitting [Equation (\ref{eq1})] and the linear approximation in the appropriate $2\omega$ range [Equation (\ref{eq2.7})], for the determination of thermal conductivity. Frequency sweeps are performed at each fixed temperature in the range of interest. The temperature-dependent thermal conductivity follows from the slope at each isotherm.

The thermal conductivity of one of our SrTiO$_3$ samples in the 50 - 300~K temperature range is shown in Figure \ref{fig:kappa}d). These data (red squares) agree very well with the values reported by Martelli \textit{et al.} \cite{martelli_prl} over a wide temperature range. Also, the room-temperature (RT) thermal conductivity of both SrTiO$_3$ and LiNbO$_3$ as determined by this method is compared to the reference values in Table \ref{tableResults}.

Table \ref{tab_values} presents a compilation of typical values and/or orders of magnitude of various parameters, based on multiple experiments we performed. 

\begin{table}[ht]
\centering
\caption{Room-temperature (RT) thermal conductivity of selected compounds from $3\omega$ experiments and available reference values. }

\begin{ruledtabular}
{\renewcommand{\arraystretch}{1.5}
\begin{tabular}{ccc}

Compound & $\kappa_{\text{300 K}}^{3\omega}$ [Wm$^{-1}$K$^{-1}$] & $\kappa_{\text{300 K}}^{\text{Ref}}$ [Wm$^{-1}$K$^{-1}$]  \\
\hline

SrTiO$_3$ & $11.3 \pm 0.4$ &  10.7\footnote{Martelli et al. (2018) \cite{martelli_prl}} \\
LiNbO$_3$ & $4.5 \pm 0.2$ & 3.3 - 5.2\footnote{ Yao et al. (2008) \cite{Yao2008}}   \\
\end{tabular}}
\end{ruledtabular}

\label{tableResults}
\end{table}

\begin{table}[ht]
\centering
\caption{Typical values and/or magnitudes of various parameters of $3\omega$ measurements used in this work.}
\label{tab_values}
\begin{ruledtabular}
{\renewcommand{\arraystretch}{1.5}
\begin{tabular}{ccc}

Parameter & Variable           & Value \\ 
\hline
H/T length & $\ell_h$      & 1 - 4 mm                   \\ 
H/T full-width & $2b_h$    & 4 - 100 $\mu$m             \\ 
H/T thickness & $t_h$      & 20 - 300 nm                 \\
H/T resistance  & $R_h$     & 10 - 500 $\Omega$\\ H/T TCR & $\beta_h$        & $\sim 2\times 10^{-4}$ K$^{-1}$            \\
Residual Resist. Ratio     & RRR                        & 1 - 3                \\
Power per unit. length     & $p_0 = P_{\text{DC}}/\ell_h$                        & 1 W/m          \\ 
Sample thickness & $t_s$              & 0.5 - 2 mm          \\  
Frequency range      & $f = 2\omega/4\pi$           & 1 - 4000 Hz         \\
WB input voltage & V$_0$                        & 1 - 5 V                \\ 
Current on H/T          & $\mathcal{O}(I_0)$              & mA        \\ 
Voltage on H/T          & $\mathcal{O}(V_{h, 1\omega})$              & mV - V         \\ 
\makecell{WB voltage \\ at 3$^{\text{rd}}$ harmonic } & $\mathcal{O}(W_{3\omega})$              & $\mu$V          \\  
\makecell{Temp. Oscillations \\ at 2$^\text{rd}$  harmonic } &$\mathcal{O}(T_{2\omega})$              & mK - K          \\  
\makecell{Operating Temp. \\ range}& $T_\infty$ & > 50 K \\
\end{tabular} }
\end{ruledtabular}
\label{tab:3omegaSummary}

\end{table}

\subsection{Error considerations}
\label{sec:error}
The following deduction of the simplified propagation of errors is based on the discussions by Hahtela \textit{et al.} \cite{Hahtela2015} and Prasad \textit{et al.} \cite{Prasad2020}. Since the third harmonic depends on the temperature oscillations via Equation (\ref{eq2}), assuming the validity of the linear regime and using the real part of Equation (\ref{eq2.7}), we can write




\begin{equation}
\centering
\label{eq_appendixD_3}
V_{h,3\omega} = -\tilde{A}\log(2\omega) + \tilde{B}
\end{equation}
where $\tilde{A} = \frac{dW_{3\omega}}{d(\log(2\omega))}\cdot \kappa$ and $\tilde{B} = \tilde{A}\cdot\left[\log(b_h^2/\alpha) - 2\xi \right]$. Taking the first derivative of Equation (\ref{eq_appendixD_3}), and combining Equations  (\ref{eq3}) and (\ref{eq:W}), we have the simplified expression for thermal conductivity as 

\begin{eqnarray}
\label{eq_appendixD_4}
    \kappa &&= \frac{I_h^3R_h^2\beta_h}{4\pi\ell_h}\left(-\diff{V_{h,3\omega}}{\log(2\omega)}\right)^{-1} \nonumber \\
&&= -\frac{1}{4} \cdot \frac{\beta_h}{\pi \ell_h  \tilde{A}} \cdot \left(\frac{V_{h,1\omega}^3}{R_h}\right) \cdot \left(\frac{R_1}{R_1 + R_h}\right)
\end{eqnarray}

To estimate the uncertainty in the measurements, the Kline \&
McClintock method \cite{Moffat1988} can be used. The general expression for the uncertainty in a thermal conductivity determination is 

\begin{eqnarray}
\label{eq_appendixD_5}
    \sigma_{\kappa}^2 = && =  \left( \frac{\partial \kappa}{\partial \beta_h} \cdot \sigma_{\beta_h} \right)^2 +
\left( \frac{\partial \kappa}{\partial \ell_h} \cdot \sigma_{\ell_h} \right)^2 \nonumber\\ &&+
\left( \frac{\partial \kappa}{\partial \tilde{A}} \cdot \sigma_{\tilde{A}} \right)^2 +
\left( \frac{\partial \kappa}{\partial V_{h,1\omega}} \cdot \sigma_{V_{h,1\omega}} \right)^2 \\ &&+
\left( \frac{\partial \kappa}{\partial R_h} \cdot \sigma_{R_h} \right)^2 +
\left( \frac{\partial \kappa}{\partial R_1} \cdot \sigma_{R_1} \right)^2\nonumber
\end{eqnarray}
where $\sigma_{x_i}$ is the uncertainty associated with the variable $x_i$. For the explicit form of each partial derivative, please see Appendix \ref{appendixA}. Finally, the relative contribution in uncertainty of variable $x_i$ to the thermal conductivity $\kappa$ is 

\begin{equation}
\label{eq_appendixD_12}
\text{``Contribution of } x_i \; \text{to} \; \sigma_\kappa\text{''} = \frac{\left( \frac{\partial \kappa}{\partial x_i} \cdot \sigma_{x_i} \right)^2}{\sigma_\kappa^2}
\end{equation}

Because of Equation (\ref{eq_appendixD_12}), to minimize the resulting uncertainty in the thermal conductivity via the $3\omega$ method, it is desirable to minimize the uncertainty of each of the terms in Equations (\ref{eq_appendixD_6}) - (\ref{eq_appendixD_11}). Table \ref{tableUncertainties} presents the uncertainties in the variables regarding the room-temperature (RT) thermal conductivity of SrTiO$_3$ (see Table \ref{tableResults}). 

\begin{table}[ht]
\centering
\caption{Values and uncertainties of variables related to room-temperature thermal conductivity of SrTiO$_3$ as determined via the $3\omega$ method. }

\begin{ruledtabular}
{\renewcommand{\arraystretch}{1.5}
\begin{tabular}{ccccc}

Variable & Value & Uncertainty & $\sigma_{x_i}/x_i$ & \makecell{Relative \\contribution \\to $\sigma_{\kappa}$}\\
\hline

$\beta_h$ & $1.9\times 10^{-4}$ K$^{-1}$ & $5.4\times10^{-5}$ K$^{-1}$ & 2.7\% & 78.0\%\\
$V_{h,1\omega}$ & 0.1835 V & $5\times10^{-4}$ V & $\sim$ 0.3\% & 7.2\%\\
$\tilde{A}$ & $-3.50\times10^{-7}$ V & $3\times10^{-9}$ V  & $\sim$ 0.8\% & 7.0\%\\
$R_h$ & 37.0 $\Omega$ & 0.2 $\Omega$ & $\sim$ 0.5\% & 5.6\%\\
$\ell_h$ & 3.70 mm  & 0.01 mm  & 0.2\% & 0.8\%\\
$R_1$ & 46.8 $\Omega$ & 0.3 $\Omega$ & $\sim$ 0.5\% & 0.5\%\\

\end{tabular}}
\end{ruledtabular}

\label{tableUncertainties}
\end{table}

Unlike the standard steady-state method, the technique has negligible errors in geometry parameters ($<$ 1\%) since the length and width of the metal line are determined to high accuracy by shadow masks and lithography. Errors associated with radiation losses are usually negligible because of the small temperature scale of the measurement \cite{cahill1990}. We notice that, consistently with literature reports\cite{Putnam2003, Chen2009}, the primary source of uncertainties in the thermal conductivity measurement comes from the calibration curve, and consequently, the TCR parameter. In particular, one may observe that the resistance of the H/T is determined upon calibration against a reference temperature sensor. However, as the temperature oscillations are measured on top of a thermal background rise\cite{Dames2013}, $R_h$ can likely be underestimated during the data analysis. Although this effect can be small, it may introduce an error offset in the thermal conductivity. 

Another critical source of error might be an elevated pressure level inside the experimental chamber. Sekimoto and co-workers \cite{Sekimoto2023} have shown that the measured thermal conductivity can be as high as a factor 2 of its true value if the pressure is not below $10^{1}$ Pa. This effect is expected to be the more pronounced the less conducting the sample is.

It is important to note that some factors impact the uncertainties of the method that cannot be accounted for through uncertainty propagation. Bad thermal design, poor heat sinking and intermediate capping layers (in the case of metallic bulk samples) may introduce significant sources of errors that are not trivial to avoid\cite{Dames2013}.

\section{Conclusions}

The 3$\omega$ method is reliable for the determination of bulk thermal conductivity, insensitive to radiation losses, and applicable in a wide temperature range 50 - 850~K, making it suitable to characterize candidate materials for several applications such as thermoelectric, thermal-coating, and batteries. Its relatively low cost \cite{Acherman2014, Langenberg2016}, as demonstrated herein, justifies the use of this technique in thermal transport studies. However, its implementation is not trivial and requires optimizing many technical, electronic, and modeling details. 

In this work, we compiled the main concepts, provided specific quantitative information to help get started with this experiment, and developed a Python-based interface to support planning a 3$\omega$ experiment on a bulk sample. 


One of the main disadvantages of the $3\omega$ method is the non-reusability of the H/T transducer. Moreover, in the case of metallic or semiconducting samples, the required electrical isolation of the micro-fabricated H/T line from the sample always introduces an extra source of thermal resistance, potentially lowering the method's accuracy \cite{Acherman2014}. The limiting factor with respect to the temperature range is the flattening of the H/T resistance curve, which impedes low temperature $\kappa(T)$ measurements. Such a constraint could be overcome by investigating alternative materials for the H/T transducer with non-vanishing low-$T$ TCR.


\begin{acknowledgments}
This study was financed by the São Paulo Research Foundation (FAPESP), Brazil, Process n. 2018/19420-3, 2022/01742-0,  2021/00989-9, 2022/14202-3, and 2018/08845-3. VM acknowledges CNPq for Grant n. 302427/2025-2. J.L.J. acknowledges CNPq Grant No. 308825/2025-0. M.S. acknowledges support of the PIBIC-CNPq program (n. 2023-1982). The authors acknowledge the financial support of  UGPN-Aucani (2018, 2020). We thank S. Romero (Laboratory for Magnetic Materials - Institute of Physics/USP) for support on Pt sputtering deposition for the $3\omega$ experiments. We also thank Marcos Santos de Souza from the Institute of Physics-USP mechanical workshop, Eduardo Thomaz Noronha, Murilo Balhester de Almeida, David Mioto Nunes de Oliveira, and João Pedro de Souza Pereira for valuable discussions and support. 
\end{acknowledgments}

\section*{Data Availability Statement}

The data supporting this study's findings are available from the corresponding author upon reasonable request.

\section*{License}
Copyright (2025) Authors. This article is distributed under a Creative Commons Attribution-NonCommercial-NoDerivs 4.0 International (CC BY-NC-ND).

\appendix

\section{Partial derivatives for uncert. propagation}\label{appendixA}

The partial derivatives to be considered in Equation (\ref{eq_appendixD_5}) are

\begin{equation}
\label{eq_appendixD_6}
\frac{\partial \kappa}{\partial \beta_h} = -\frac{1}{4} \cdot \frac{1}{\pi  \ell_h  \tilde{A}} \cdot \left(\frac{V_{h,1w}^3}{R_h}\right) \cdot \left(\frac{R_1}{R_1 + R_h}\right)
\end{equation}

\begin{equation}
\label{eq_appendixD_7}
\frac{\partial \kappa}{\partial \ell_h} =- \frac{1}{4} \cdot \frac{\beta_h}{\pi \ell_h^2  \tilde{A}} \cdot \left(\frac{V_{h,1w}^3}{R_h}\right) \cdot \left(\frac{R_1}{R_1 + R_h}\right)
\end{equation}

\begin{eqnarray}
\label{eq_appendixD_8}
\frac{\partial \kappa}{\partial \tilde{A}} = -\frac{1}{4} \cdot \frac{\beta_h}{\pi  \ell_h  \tilde{A}^2} \cdot \left(\frac{V_{h,1w}^3}{R_h}\right) \cdot \left(\frac{R_1}{R_1 + R_h}\right) 
\end{eqnarray}

\begin{eqnarray}
\label{eq_appendixD_9}
\frac{\partial \kappa}{\partial V_{h,1w}} = -\frac{1}{4} \cdot \frac{\beta_h}{\pi \cdot \ell_h \cdot \tilde{A}} \cdot \left(\frac{3 V_{h,1w}^2}{R_h}\right) \cdot \left(\frac{R_1}{R_1 + R_h}\right)
\end{eqnarray}

\begin{eqnarray}
\label{eq_appendixD_10}
\frac{\partial \kappa}{\partial R_h} &&= -\frac{1}{4} \cdot \frac{\beta_h}{\pi \cdot \ell_h \cdot \tilde{A}} \cdot \left( \frac{V_{h,1w}^3}{R_h^2} \right) \cdot \left( \frac{R_1}{R_1 + R_h} \right) \nonumber \\&&- \frac{1}{4} \cdot \frac{\beta_h}{\pi \cdot \ell_h \cdot \tilde{A}} \cdot \left( \frac{V_{h,1w}^3}{R_h} \right) \cdot \left( \frac{R_1}{(R_1 + R_h)^2} \right)
\end{eqnarray}

\begin{eqnarray}
\label{eq_appendixD_11}
\frac{\partial \kappa}{\partial R_1} &&= \frac{1}{4} \cdot \frac{\beta_h}{\pi \cdot \ell_h \cdot \tilde{A}} \cdot \left( \frac{V_{h,1w}^3}{R_h} \right) \cdot \left( \frac{1}{R_1 + R_h} \right) \nonumber \\&&- \frac{1}{4} \cdot \frac{\beta_h}{\pi \cdot \ell_h \cdot \tilde{A}} \cdot \left( \frac{V_{h,1w}^3}{R_h} \right) \cdot \left( \frac{R_1}{(R_1 + R_h)^2} \right)
\end{eqnarray}


%

\end{document}